\documentclass[letterpaper, 10 pt, conference]{ieeeconf}  

\IEEEoverridecommandlockouts                              

\usepackage{cite}
\usepackage{amsmath,amssymb,amsfonts}
\usepackage{graphicx}
\usepackage{textcomp}
\usepackage{xcolor}
\graphicspath{{figures/}}

\usepackage{siunitx} 
\usepackage{algorithm}
\usepackage{algpseudocode}
\usepackage[hidelinks]{hyperref}
\usepackage[capitalise]{cleveref}

\usepackage{newfloat}
\usepackage{listings}
\usepackage{comment}

\everypar{\looseness=-1}

\floatstyle{ruled}
\newfloat{listing}{tb}{lst}{}
\floatname{listing}{Listing}
\newcommand\m[1]{\begin{bmatrix}#1\end{bmatrix}}

\newcommand{\tp}{\mathsf{T}}

\title{\LARGE \bf
Model Predictive Planning: Trajectory Planning in\\ Obstruction-Dense Environments for Low-Agility Aircraft
}

\author{Matthew T. Wallace$^{1}$,  Brett Streetman$^{2}$, and Laurent Lessard$^{3}$
\thanks{This work was supported by a Draper Scholarship}
\thanks{$^{1}$M.T. Wallace is a Draper Scholar and is with the dept. of Electrical and Computer Engineering, Northeastern University, Boston, MA 02115, USA}%
\thanks{$^{2}$B. Streetman is with The Charles Stark Draper Laboratory, Cambridge, MA 02139, USA}%
\thanks{$^{3}$L. Lessard is with the dept. of Mechanical and Industrial Engineering, Northeastern University, Boston, MA 02115, USA\newline {\tt\small l.lessard@northeastern.edu}}
}

\usepackage{amsmath}
\begin{document}
\maketitle

\begin{abstract}
We present Model Predictive Planning (MPP), a trajectory planner for low-agility vehicles such as a fixed-wing aircraft to navigate obstacle-laden environments.  MPP consists of (1) a multi-path planning procedure that identifies candidate paths, (2) a raytracing procedure that generates linear constraints around these paths to enforce obstacle avoidance, and (3) a convex quadratic program that finds a feasible trajectory within these constraints if one exists. Low-agility aircraft cannot track arbitrary paths, so refining a given path into a trajectory that respects the vehicle's limited maneuverability and avoids obstacles often leads to an infeasible optimization problem. The critical feature of MPP is that it efficiently considers multiple candidate paths during the refinement process, thereby greatly increasing the chance of finding a feasible and trackable trajectory. We demonstrate the effectiveness of MPP on a longitudinal aircraft model.
\end{abstract}

\section{Introduction}

We consider the problem of autonomous navigation and obstacle avoidance for a low-agility vehicle, such as a fixed-wing aircraft. In this context, it is important to distinguish between a \emph{path} and a \emph{trajectory}.

A \emph{path} describes \emph{where} the vehicle moves in space as it travels from its starting position to its final destination. In this paper, paths are parameterized as a sequence of spatial coordinates and include the straight lines connecting them. Paths are only concerned with the geometric course that the vehicle follows and do not specify the timing, speed, or orientation of the vehicle as it moves along. A \emph{feasible path} is one that connects the starting position and the final destination and does not intersect any obstacles. 
Methods like Rapidly-exploring Random Trees* (RRT*) iteratively construct paths that are guaranteed to converge to the global shortest feasible path and can handle many geometric constraints \cite{karaman2011sampling}.  

A \emph{trajectory} describes \emph{how} the vehicle moves in space. It specifies how the full state of the vehicle evolves as a function of time. A \emph{feasible trajectory} is one that connects the starting position and the final destination, satisfies the equations of motion of the vehicle, and does not collide with any obstacles. For high-agility vehicles such as quadrotors, we can navigate around obstacles by generating a feasible path and using a reference-tracking controller to track the path. However, this approach may fail with low-agility vehicles since they cannot change directions abruptly and may collide with nearby obstacles as they attempt to follow the prescribed path.

\begin{figure}[ht]
\centering
\includegraphics[width=\linewidth]{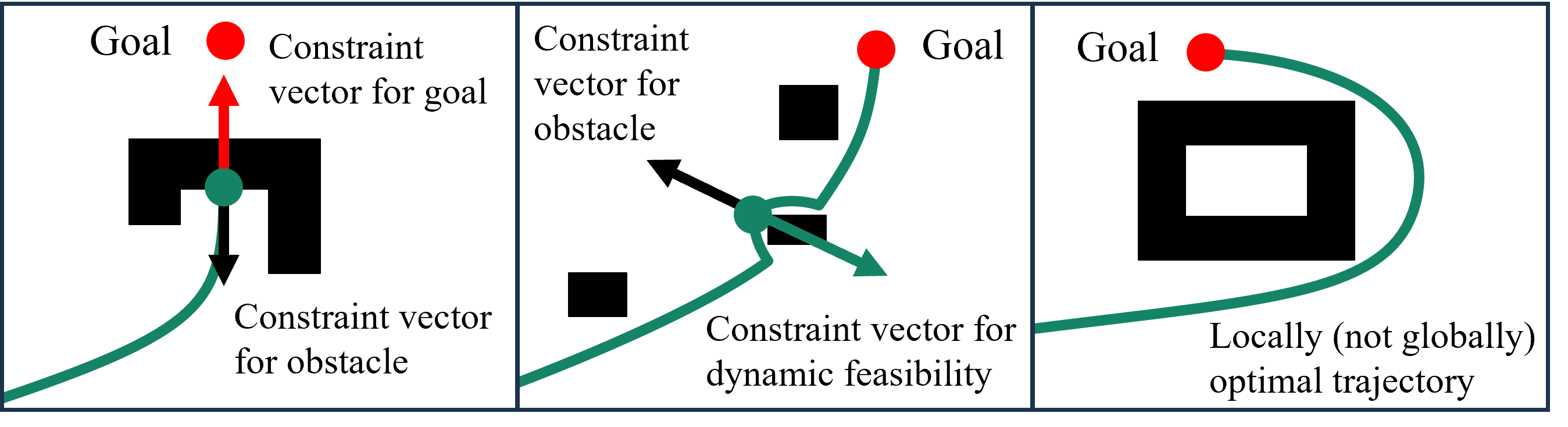}
\caption{A visualization of some challenges of trajectory optimization around obstacles.  \textbf{Left:} the trajectory is caught in a flytrap; the vector pointing to the goal directly opposes the obstacle constraint vector. \textbf{Middle:} the constraint vector for the obstacle directly opposes the dynamic constraint. These two cases result in optimizers returning no solution. \textbf{Right:} optimizer chooses the long way around the obstacle and gets stuck in a local minimum.\looseness=-1}
\label{fig:OptimizationChallenges}
\end{figure}

In general, trajectory optimization amounts to solving a nonlinear optimal control problem \cite{rao2014trajectory} and is far more difficult than finding a feasible path. Examples of tools that can perform offline trajectory optimization include PROPT \cite{rutquist2010propt} and GPOPS-II \cite{patterson2014gpops}. As with most nonlinear optimization solvers, these methods find locally optimal solutions and are highly dependent on having good initial guesses. Including obstacle avoidance can be problematic because it can create more local minima and further increase the reliance on initial guesses.
For example, if an obstacle were to split the space in half so that a vehicle could go either left or right, and only one of these two choices leads to a feasible trajectory, the wrong initial guess would cause the solver to fail.  See \cref{fig:OptimizationChallenges} for an illustration of some difficult scenarios for trajectory optimization.
The strong dependence on initial guesses persists when performing trajectory optimization in an in-the-loop fashion using a traditional Model Predictive Control (MPC) architecture.

To address the problem of dependence on initial guess, we propose a new approach for in-the-loop trajectory optimization and obstacle avoidance called Model Predictive Planning (MPP). The main components of MPP are as follows.
\begin{enumerate}
    \item A multi-path planner based on RRT*-AR \cite{6631133} that generates several candidate paths.
    \item A raytracing procedure that, for a given candidate path, generates linear constraints that enforce obstacle avoidance.\looseness=-1
    \item A convex quadratic program that uses the linear constraints to refine each candidate path into a feasible trajectory if one exists. We then select the best feasible trajectory among those that were found.
\end{enumerate}

MPP works in an in-the-loop fashion similar to MPC. However, since MPP does not require solving nonlinear optimization problems, it can more efficiently evaluate whether a given candidate path can lead to a feasible trajectory. Therefore, the available computational effort can be spent evaluating numerous candidate paths.  MPP is analogous to a breadth-first search (search many paths) rather than the depth-first approach employed in trajectory optimization (use a single path to initialize a nonlinear solver). 

A number of motion planning approaches employ multiple trajectories.  \emph{Motion primitive planning} is one such method that has been demonstrated on quadrotors \cite{icra2019}, robotic manipulators \cite{shyam2019improving}, and fixed wing drones \cite{bulka2021control}.  This method reliably avoids obstacles, but can produce inefficient trajectories since it involves selecting among a limited subset of possible control actions.  Many trajectories are also used in \emph{stochastic optimization} methods such as Stochastic Trajectory Optimization for Motion Planning (STOMP) \cite{kalakrishnan2011stomp} and Model Predictive Path Integral control (MPPI) \cite{williams2017information}.
These methods refine a baseline trajectory by sampling nearby trajectories, which differs from MPP in that they cannot explore topologically distinct trajectories.
Another relevant work is the Fast and Safe Trajectory Planner (FASTER) \cite{tordesillas2021faster}, which plans two trajectories (aggressive and safe) for a quadrotor navigating in an unknown environment.

A few works explicitly address using multiple trajectories to avoid problems with local minima.  Zhou et al. \cite{zhou2020robust} use a topological search to generate multiple paths for a quadrotor.  These paths are refined into B-spline trajectories.  Our work differs in our choice of path planner, an RRT* variant rather than a topological planner, and that we use full trajectory optimization rather than path smoothing.  Using the RRT* variant could potentially generate more diverse paths in a field with many obstacles as the topological search was limited to only 5 paths due to computational limitations.  Using full trajectory optimization allows our method to apply to low-agility systems like fixed-wing aircraft.

\emph{Differential Dynamic Programming} (DDP) \cite{jacobson1968new} is a popular numerical approach for solving the Hamilton--Jacobi--Bellman equation via alternating shooting and back-propagation.  Maximum Entropy DDP \cite{so2022maximum}, a recent adaptation of DDP, is able to find multiple paths through an obstacle field.  The strategy has the benefit of combining the path planning and optimization steps used in our algorithm, but the optimization problem proposed lacks the guarantees of convex optimization and additionally cannot handle constraints explicitly, but instead uses soft constraints.

Topological-driven Model Predictive Control (T-MPC) \cite{de2024topology} uses a \emph{probabilistic roadmap} and a depth-first search to generate paths through a field of moving obstacles and then solves local nonconvex optimization problems to refine each path.  Our approach uses convex optimization, which is generally more computationally efficient and reliable.

None of the aforementioned works explicitly treat the problem of obstacle avoidance for fixed-wing aircraft. Some works on fixed-wing aircraft include \cite{lin2020fast,bulka2019high}, but these focus on modifying an existing flight path to avoid a detected obstacle rather than reoptimizing the full trajectory.

\cref{sec:sys_mode} presents our system architecture, \cref{sec:algorithm} develops the path planner and trajectory optimizer algorithms, \cref{sec:results} demonstrates MPP on a longitudinal flight model, and \cref{sec:conclusion} states conclusions and future work. 

\begin{figure*}[ht]
  \centering
  \includegraphics{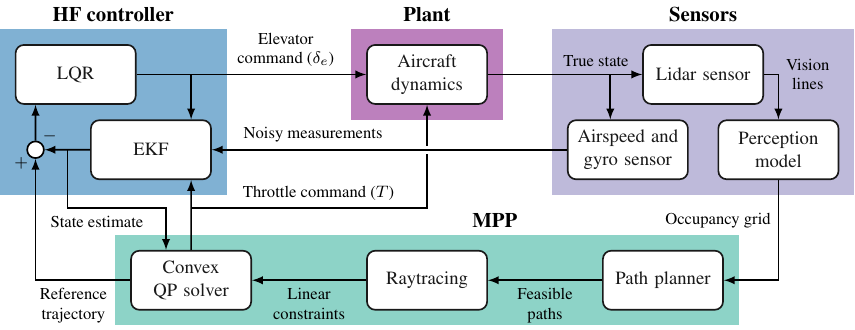}
  \caption{The system architecture, which consists of four components: the plant, sensor models, high-frequency (HF) controller, and model predictive planner (MPP). 
 The HF controller regulates the system and also provides a high-frequency (\qty{100}{Hz}) estimate of the airspeed and angles for trajectory optimization.  MPP begins with planning a path from the occupancy grid, which is then used to determine constraints on an optimization problem through raytracing, and finally a convex quadratic program (QP) is solved to find the trajectory.}
  \label{fig:arc}
\end{figure*}

\section{System Model} \label{sec:sys_mode}
Our system architecture, shown in \cref{fig:arc}, is inspired by Ryll et al.~\cite{icra2019}.  
We discuss our plant and dynamics in \cref{subsec_1}.  
The high-frequency (HF) controller operates on information provided by the airspeed and gyro sensors to regulate the vehicle, which is discussed in \cref{subsec_2}.  The lidar sensor model provides information about nearby obstacles, which is refined into an occupancy grid in the perception model, discussed in \cref{subsec_3}.
\subsection{Aircraft Dynamics Model} \label{subsec_1}
\begin{figure}[ht]
\centering
\includegraphics{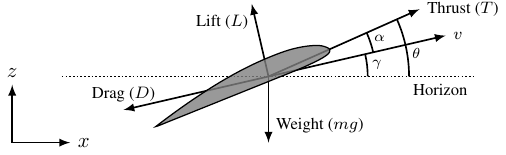}
\caption{Free body diagram of forces acting on the system.
\vspace*{-2mm}}
\label{fig:cords}
\end{figure}

We use the following longitudinal flight dynamics model. The states are downrange distance, vertical height, airspeed, pitch, pitch rate, and flight path angle $(x, z, v, \theta, \dot{\theta}, \gamma)$. The inputs are thrust and elevator deflection $(T,\delta_e)$ \cite{Stengel}.  The equations of motion are
\begin{align*}
    \dot{x} &= v\cos{\gamma},  &
    \dot{v} &= -\tfrac{D}{m}-g\sin{\gamma} + \tfrac{T}{m}\cos{\alpha},\\
    \dot{z} &= v \sin{\gamma}, &
    \dot{\gamma} &= \tfrac{L}{mv}-\tfrac{1}{v}\cos{\gamma}- \tfrac{T}{mv}\sin{\alpha} ,\\
    \ddot{\theta} &= I_{yy}^{-1}M( \delta_e).
\end{align*}
The angle of attack ($\alpha := \theta-\gamma$) is defined as the angle between the wings and oncoming airflow. Drag ($D$), lift ($L$), and pitching moment ($M$) are aerodynamic terms that depend on the states and inputs (see Appendix). Mass ($m$), moment of inertia ($I_{yy}$), and gravity ($g$) are fixed parameters.   We modeled our system as having airspeed and gyro sensors, which provide measurements of airspeed and pitch, respectively, each with Gaussian errors.
See \cref{fig:cords} for a diagram of the modeled forces and angles. 

\subsection{High-Frequency Controller}\label{subsec_2}
\quad The high-frequency (HF) controller consists of a Linear Quadratic Regulator (LQR) and an Extended Kalman Filter (EKF).  The LQR regulates the flight path angle response of the system using elevator action. We linearize around straight and level flight conditions.  The HF controller is responsible for tracking the desired pitch angle determined by MPP.  The HF controller states are $(v,\theta,\dot\theta,\gamma)$, input is $\delta_e$, and the associated $Q$ and $R$ matrices for the LQR controller were chosen as
$Q = \mathrm{diag}(1,1,0,1000)$ and $R=\tfrac{1}{2}$.
We used $Q_{33}=0$ because $\dot\theta$ is not directly measured and its estimate is noisy. We used $Q_{44} = 1000$ because the regulation of flight path angle is critical for obstacle avoidance.  This HF control loop does not attempt to correct for positional error, which is only considered in the trajectory planner.  We used the continuous time LQR solver from the Python Control Systems Library \cite{python-control2021} to calculate our LQR gains.  

The EKF estimates the true states based on pitch and airspeed indicator measurements.  The estimated state is used both in the HF controller and trajectory planner.  

\subsection{Perception model}\label{subsec_3}
\begin{figure}[ht]
\centering
\includegraphics{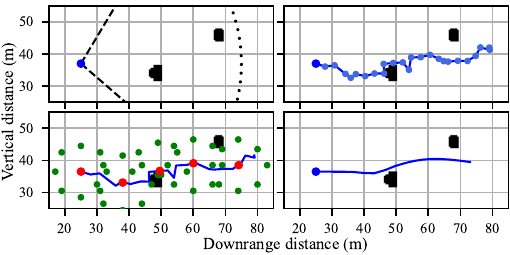}
\caption{The MPP trajectory creation process.
\textbf{Top-left:} Lidar sweep used to create an occupancy grid.
\textbf{Top-right:} Feasible path generated from the occupancy grid and RRT*-AR.
\textbf{Bottom-left:} Raytracing around the path. For clarity, only 1/100 of the collocation points are shown (in red); see \cref{fig:searchFig} for more details.
\textbf{Bottom-right:} Trajectory produced from the raytracing by solving a convex QP.}
\label{fig:TrajGen1}
\end{figure}

We modeled our perception system using a generated obstacle field and simulated lidar. The model identified objects at a particular angle and range with an error proportional to the range. These ray measurements are converted into an occupancy grid: the $(x,z)$ coordinates where lidar rays intersect obstacles in the field of view. The first pane of \cref{fig:TrajGen1} shows the output occupancy grid from the perception model.

\section{MPP Algorithm}\label{sec:algorithm}

Pseudocode for the MPP algorithm is given in \cref{alg:MPP}.  The inputs to this method are an estimated dynamic state and occupancy grid, supplied by the EKF and perception model respectively.  The path planner, 
a variation on RRT discussed in \cref{sec:pathplanning}, takes in the occupancy grid and identifies paths through this space towards the goal state of 54 meters directly ahead. The raytracing procedure converts these paths and the occupancy grid to convex optimization constraints and is discussed in \cref{sec:raytracing}.  The resulting optimization problem is solved in CVXOPT, which is discussed in \cref{sec:trajopt}.  \cref{fig:TrajGen1}
 shows the algorithm run with a single path.  Because a path determined by a path planning may not correspond to a feasible trajectory, we optimize several paths rather than a single one.  This approach is shown in \cref{fig:multiplePaths}.\looseness=-1
\begin{algorithm}
\caption{Model Predictive Planning}\label{alg:MPP}
\begin{algorithmic}
\State $x_0 \leftarrow$ EKF
\State OccupancyGrid $\leftarrow$ PerceptionModel
\State Paths $\leftarrow$ RRT$^*$-AR(OccupancyGrid)
\State Trajectory$^*$ 
\For{Path in Paths}
    \State PositionConstraints $\leftarrow$ Raytrace(Path, OG)
    \State Trajectory $\leftarrow$ CVX($x_0$, PositionConstraints)
    \State Trajectory$^*$ $\leftarrow$ mincost(Trajectory, Trajectory$^*$)
\EndFor
\end{algorithmic}
\end{algorithm}

\subsection{Path Planning}\label{sec:pathplanning}
\begin{figure}[ht]
\centering
\includegraphics{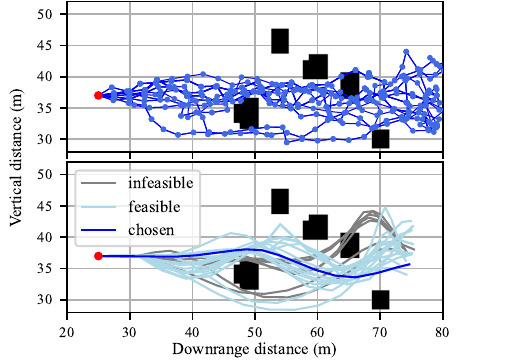}

\caption{Visualization of the multi-trajectory process.  The process is similar to \cref{fig:TrajGen1}, but many paths are planned instead of just one.  Each candidate path is refined into a trajectory. Infeasible trajectories are rejected and the best feasible trajectory is chosen.}
\label{fig:multiplePaths}
\end{figure}

Path Planning is carried out with the RRT variant RRT*-AR, which differs from RRT* in that it adds a cost penalty to close nodes sharing the same parent node \cite{6631133}.  Like standard RRT, the RRT*-AR outputs obstacle-free paths.  This approach produces several suboptimal alternate routes to the goal.  We found running RRT*-AR a single time provided insufficient dispersion of paths, but that sufficient dispersion was achieved by rerunning RRT*-AR with a low number of samples each time.  The results of a single run are shown in the second pane of \cref{fig:TrajGen1} and for 50 sampled paths in the first pane of \cref{fig:multiplePaths}.

\begin{figure}[ht]
\centering
\includegraphics[width=\linewidth]{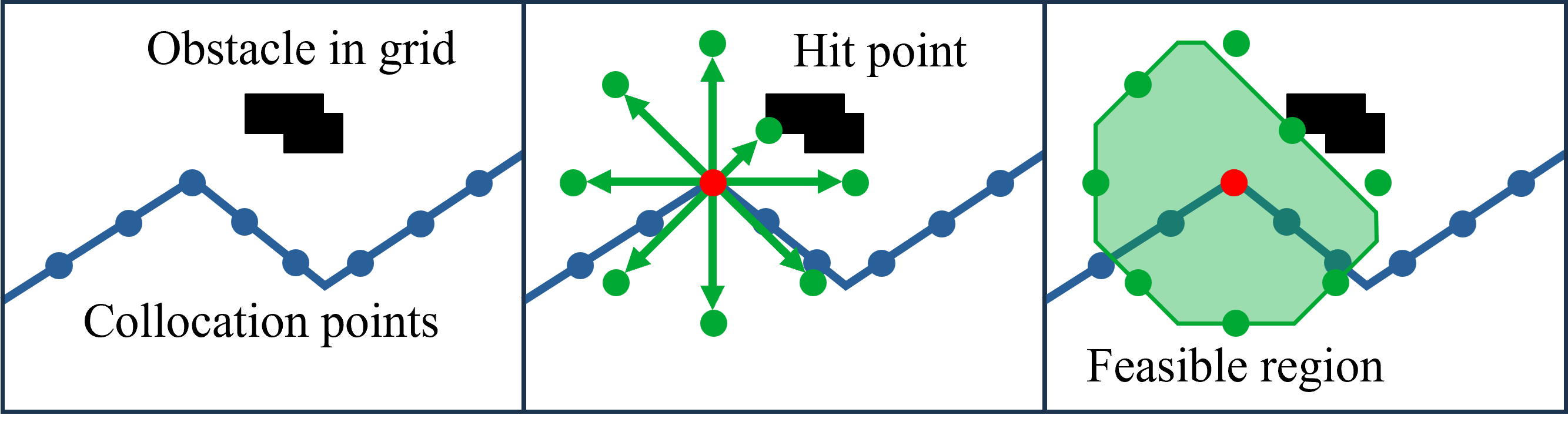}
\caption{Visualization of raytracing (see bottom-left pane in \cref{fig:TrajGen1}).
\textbf{Left}: \emph{Collocation points} are placed along the path (equally spaced in time).
\textbf{Middle}: Rays (in green) are drawn in different directions from a collocation point (in red) until they reach an obstacle or travel a given distance.
\textbf{Right}: Each ray is the normal vector to a linear constraint and these constraints define a convex region.
This process is repeated with all collocation points. Each point is allowed to move within its prescribed convex region subject to satisfying vehicle dynamics in order to produce an improved path.\looseness=-1}
\label{fig:searchFig}
\end{figure}
\subsection{Raytracing}\label{sec:raytracing}

The raytracing procedure determines a feasible region around a path in which to search for a trajectory.  While the optimization problem over the whole space is not convex, this local search problem is convex and thus much more practical to solve online.  Our approach is inspired by the algorithm from Meerpohl et~al. \cite{MEERPOHL2019368}.  

Beginning from points on the path, we check the occupancy grid (stored as a binary matrix) along 8 directions, returning when an occupied point is encountered.  This process is visualized in \cref{fig:searchFig}.  Hit points are shown in the third pane of \cref{fig:TrajGen1}.  Hit points are translated to their position in a linearized straight and level vehicle frame and then an inner product is taken with the search direction, resulting in a linear constraint.  Constraints are padded slightly to account for rounding error and provide margin for tracking.\looseness=-1

\begin{figure*}[ht]
  \centering
  \includegraphics{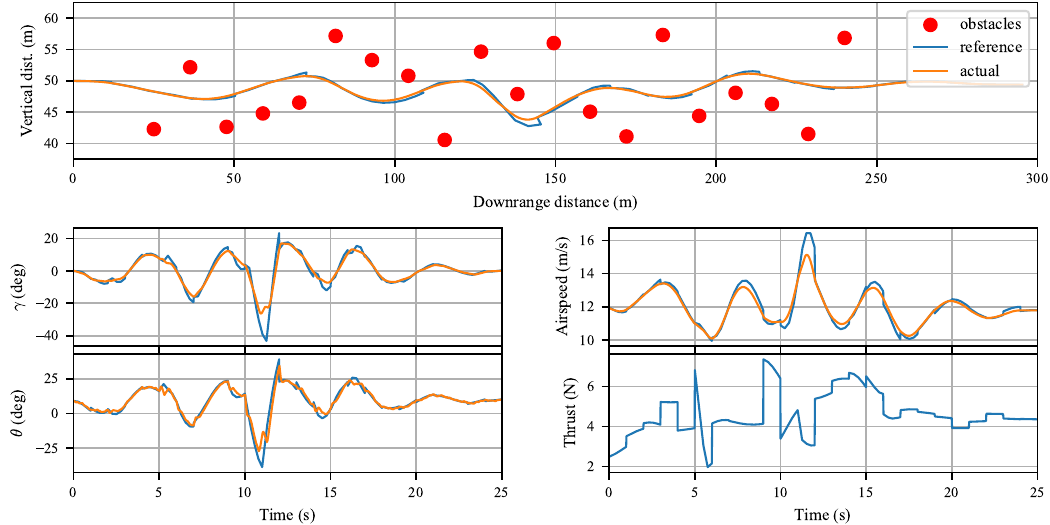}
  \vspace*{-3mm}
  \caption{A sample run of MPP with 20 obstacles and 25 paths per iteration.  The reference trajectory is the output of MPP, which is then tracked by the HF controller in closed loop. The reference trajectory (blue) is jagged because it is re-planned every second (with a 4.5 second horizon), at which point it snaps to the current estimate of the system state. The true state is shown in orange.\vspace*{-3mm}}
  \label{fig:MPPFull}
\end{figure*}

\subsection{Trajectory Optimization} \label{sec:trajopt}

The trajectory optimization problem we solve is a convex quadratic program (QP). For an introduction on convex optimization for trajectory generation, see Malyuta et al.~\cite{9905530}.  This method is a form of direct optimization, solving for a perturbation off an initial guess rather than exploiting the Pontryagin principle.  Since our dynamics are near-linear, we were able to obtain good results with a single perturbation step using straight and level flight as an initial guess without the need for iterative linearization as is often necessary in highly nonlinear cases.
We used CVXOPT \cite{cvxopt} to solve the QP. The optimization variable is $\bar{x}$, the perturbation of states and controls from straight and level flight over the forward time horizon. The problem parameters are the: quadratic cost $\bar{Q}$, equality constraint from the dynamics $\bar{A}$, vector of zeros in the equality constraint $\bar{b} = 0$, inequality constraint matrix to capture the search directions $\bar{G}$, and corresponding vector from raytracing $\bar{h}$.  The optimization problem is\looseness=-1
\begin{equation*}
\min\; \bar{x}^{\tp} \bar{Q}\bar{x} \quad \text{s.t.} \quad  \bar{A}\bar{x} = \bar{b}, \,\,  \,\, \bar{G} \bar{x} \leq \bar{h}.
\end{equation*}

The system is linearized around straight and level flight.  $x_0$ is the difference from those conditions at the current time.   Those dynamics are captured by the equality constraints
\begin{align*}
\bar{x} =\! \m{\delta x_0\\ \delta u_0\\ \vdots \\ \delta x_{n-1}\\\delta u_{n-1} \\\delta x_n}\!,\,\, 
\bar{A} =\! \addtolength{\arraycolsep}{-1pt}\m{I &0&0&\cdots&0\\A &B &-I&\cdots&0\\ \vdots &\ddots&\ddots&\ddots& \vdots \\0&\cdots& A&B &-I}\!,\,\,
\bar{b} =\! \m{x_0\\ 0\\ \vdots\\0}\!,
\end{align*}
where $A$ and $B$ are the linearizations of the dynamics with respect to the entire state and control input, respectively.  The feasible region found from raytracing is encoded as inequality constraints.  The unit length search directions $g_1,\dots,g_p$ are identical for each point and determine the inequality constraint matrix $\bar{G}$: 
\begin{align*}
G = \m{g_1 & \cdots & g_p}^{\tp},\qquad
\bar{G} = \m{I_n \otimes \m{G  & 0_{p \times 2}} & \\ & G}.
\end{align*}
For the experiments in this paper, we used $p=8$.

The raytracing vector is calculated by taking the inner product of the vector to the ``hit'' point ($o$) from the collocation point and the search direction.  The inner product of the new reference coordinate and the search direction must be less than this value of the coordinate to be inside the open region. Therefore, we have

\[ 
h_i = \m{g_1^{\tp} o_{i1} & \cdots & g_p^{\tp} o_{ip}}^\tp,\qquad
\bar{h} = \m{h_0^\tp & \cdots & h_n^\tp}^\tp.
\]

An additional inequality constraint is used to ensure the thrust $T$ is nonnegative.

The cost function consists of a penalty on the control and the terminal state's deviations from straight and level flight values.  We chose the cost penalty matrix to be
\begin{equation*}
\bar{Q} = \m{ I_n \otimes \m{0_{6\times6} & 0_{6\times 2} \\ 0_{2 \times 6}& C_{\textup{ctrl}}} & \\ & 100\cdot I_{6} }, \quad
C_{\textup{ctrl}} = \m{1 & 0\\0&3}.
\end{equation*}
We chose $C_{\textup{ctrl}}$ and the final weight of $100$ empirically; using a greater penalty for thrust than elevator produced smoother trajectories. Obstacle avoidance is enforced solely in the constraints and does not appear in the cost function.

The problem is convex by construction since $\bar Q$ is positive semidefinite. Intuitively, the linear constraints provided by the raytracing step yield a convex inner approximation of the original non-convex occupancy grid.

\subsection{Dynamic modifications}
Trajectory optimization is conducted using a time resolution of 0.25 seconds. This resolution is inadequate for accurately capturing the rotational dynamics exhibited by the system. To address this limitation, we introduced adjustments to the equality constraints associated with the rotational states. The differential equations are transformed into algebraic ones, constraining these states to their steady-state values after oscillations have decayed. In scenarios involving a single variable, this transformation takes on the following structure:
$
x_{t+1} = Ax_{t} + B u_t \implies (I-A) x_{t+1} = B u_{t}.
$
In the optimization equations, this takes the form 
\begin{equation*}
\Bar{A}_i = \m{0&\cdots&0 & B & (I-A) &0&\cdots&0}.
\end{equation*}

In addition to allowing the planner to run at a lower resolution, this procedure prevents negative interactions between the HF controller and the path planner.\looseness=-1
\section{Results}\label{sec:results}

\cref{fig:MPPFull} shows a typical successful run of MPP using 25 path candidates through a field of \num{20} 1-\unit{m} radius obstacles.  The vehicle's vision is limited to \qty{45}{\m} ahead and it cruises at \qty{12}{\m/\s}.  The trajectory is re-planned every \qty{4.5}{\s} (\qty{54}{m}).  The reference trajectory and open loop commands are calculated at \qty{1}{\Hz}, while the controller and dynamics resolve at \qty{100}{\Hz}.
This simulation demonstrates that MPP can effectively plan feasible trajectories through highly obstructed environments when implemented in a receding horizon fashion.\looseness=-1

\begin{figure}[bht]
\centering
\includegraphics{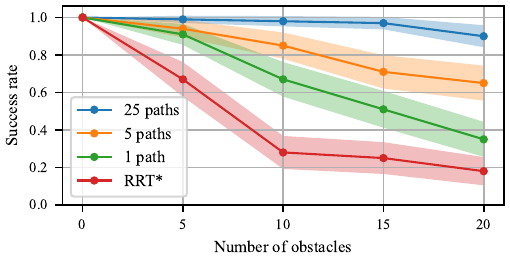}
\vspace*{-6mm}
\caption{MPP success rate based on number of paths and number of obstacles.  With a single path, using MPP to refine the path into a feasible trajectory improves RRT*'s success rate from 19\% to 35\% in a cluttered environment (20 obstacles).  The improvement increases to 90\% when using 25 paths.  Using 1 path is representative of traditional MPC. Each dot is the average of 100 trials. Shaded region indicates 95\% confidence interval.}
\label{fig:results}
\end{figure}

\cref{fig:results} shows the aggregate performance of many runs over different randomly generated obstacle fields with different numbers of obstacles. For each run, ``failure'' occurs if either an extreme state is reached (pitch exceeds \ang{60} or flight path angle exceeds \ang{45}) or there is a collision with an obstacle. For comparison, we also plot the performance of RRT* without any refinement. In other words, the feedback controller attempts to directly track the path provided by RRT*. This approach often fails because such paths may make abrupt changes that the vehicle cannot track. Refining the path using MPP yields improvement, but there are still many failures due to the fact that the initial candidate path may not lead to a feasible trajectory. As we include more candidate paths, the success rate increases. The tradeoff for this higher success rate is greater computational load, which is proportional to the number of paths.

Code that reproduces all the simulations and figures in this paper is available at
{\small\url{https://github.com/QCGroup/mpp}}.

\section{Conclusion}\label{sec:conclusion}
MPP is a modification of traditional MPC to plan trajectories around obstacles and is well-suited for low-agility vehicles.  The key feature of MPP is to use many candidate paths, thereby improving the chance of finding a feasible trajectory.\looseness=-1

We demonstrated MPP on a longitudinal flight model and found that  MPP performance improves markedly as we increase the number of candidate paths, particularly when there are many obstacles. Moreover, MPP performs better than using a path-planner alone due to the low agility of the vehicle.\looseness=-1

There are several possible avenues for improvement of MPP. First, the RRT*-AR path planner often produces topologically equivalent paths, which leads to redundant computation. If the path planner could be modified to produces topologically distinct paths, it would lead to a substantial performance improvement.
Second, MPP could potentially benefit from multiple rounds of linearization and refinement for more strongly nonlinear systems.
Finally, raytracing could be accomplished with a faster parallel implementation, which would solve the primary computational bottleneck of the algorithm.\looseness=-1
 
\bibliographystyle{IEEEtran}
\bibliography{V4}

\appendix

Lift, drag, and pitching moment are modeled as functions of coefficients ($C_L$, $C_D$, and $C_M$ respectively) that depend on velocity ($v$), elevator angle ($\delta_e$), angle of attack ($\alpha := \theta - \gamma$), and various parameters.
\begin{gather*}
        L = \tfrac{1}{2} \rho v^2 S C_L, \quad
        D = \tfrac{1}{2} \rho v^2 S C_D, \quad
        M = \tfrac{1}{2} \rho v^2 S c C_M, \\
        C_L =  C_{L0} + C_{L\alpha}\alpha, \qquad
         C_D = C_{D0} + KC_L^2, \\
        C_M =  C_{M0} + C_{M\alpha}\alpha + C_{M\dot{\alpha}}\dot{\alpha} + C_{M\delta_e}\delta_e.
\end{gather*}
The parameters used in our simulations are based on estimates for a small hobby remote-controlled aircraft:
\begin{align*}
m &=  \qty{3.2}{\kg},
& C_{L\alpha} &= \qty{5.73}{\per\radian},
& C_{L0} &= 0.5,\\
K &= 0.05, &
C_{M\alpha} &=\qty{-8.02}{\per\radian},
& C_{M0} &= 0.5, \\
S &= \qty{0.25}{\m\squared}, &
C_{M\dot{\alpha}} &= \qty{-0.46}{\s\per\radian}, &
C_{M\delta_e} &= 0.2,\\
c &= \qty{0.13}{\m}, &
\rho &= \qty{1.225}{\kg\per\m\cubed}, &
C_{D0} &= 0.1,\\
g &= \qty{9.81}{\m\per\s\squared}, &
I_{yy} &= \qty{0.17}{\kg\m\squared}.
\end{align*}

\end{document}